\documentclass[CTP,preprint]{revtex4}%
\usepackage{amsfonts}
\usepackage{amsmath}
\usepackage{amssymb}
\usepackage{graphicx}%
\begin{document}
\title{A divergent volume for black holes calls for no \textquotedblleft
firewall\textquotedblright\ }
\author{Baocheng Zhang}
\email{zhangbc.zhang@yahoo.com}
\affiliation{School of Mathematics and Physics, China University of Geosciences, Wuhan
430074, China}
\author{Li You}
\email{lyou@tsinghua.edu.cn}
\affiliation{State Key Laboratory of Low Dimensional Quantum Physics, Department of
Physics, Tsinghua University, Beijing 100084, China}
\affiliation{Frontier Science Center for Quantum Information, Beijing, China}
\keywords{volume, entropy, black hole thermodynamics}
\pacs{04.70.Dy, 04.70.-s, 04.62.+v, }

\begin{abstract}
The presumption that Hawking radiations are thermally distributed can be
considered to result from their entanglement with the internal degrees of
freedom for a black hole. This leads to the \textquotedblleft
firewall\textquotedblright\ paradox if unitary evolution continues into Page's
time when a black hole evaporates away half of its initial entropy. However,
if the interior of a black hole houses sufficient degrees of freedom to
maintain entanglement with the outside at all times, unitarity can be
preserved during the complete radiation process and no firewall will be
required. This paper proposes a scenario that rescinds firewall by introducing
the concept of volume for a black hole. Based on the operational definition by
Christodoulou and Rovelli \cite{cr15}, we show that the volume and its
associated entropy for a collapsed black hole diverges if the final
evaporation stage is treated using noncommutative space. This implicates the
interior of a black hole possesses adequate space to store information for a
black hole of any mass, like the inside of a \textquotedblleft magician's
bag\textquotedblright.

\end{abstract}
\maketitle

\section{Introduction}

Since Hawking radiation was discovered \cite{swh74,swh75}, its reconciliation
with quantum theory has been continuously contested. Irrespective of what
initial state a black hole starts with before collapsing, a thermal state of
Hawking radiations in the end is inconsistent with unitarity required for
quantum evolution. This controversy is now widely known as \textquotedblleft
the black hole information loss paradox\textquotedblright\cite{swh76}. Various
approaches were proposed for its resolution over the past few decades (see
\cite{dh16,jp16,uw17,as17} for recent reviews), although none has been
completely successful. With each failed attempt, the situation concerning the
very existence of this paradox has become more serious. Especially after the
possibility that information about infallen matter could hide into
correlations between Hawking radiations and the internal states of a black
hole is ruled out \cite{bp07}, the choice for a resolution becomes very
limited. If information were not lost, it must be either taken out by the
Hawking radiations or stored inside a black hole.

According to Page \cite{dnp93}, information is leaked out through Hawking
radiations, whose thermal spectrum (upon tracing out everything else),
reflects their entanglement with the degrees of freedom inside a black hole.
After Page's time, when half of the initial black hole entropy is evaporated,
the interior of the remaining black hole would not have sufficient degrees of
freedom to entangle with emitted radiations. Thus, if unitarity were preserved
while information escaped out from a black hole, initial entanglement between
the interior and the exterior of a black hole must be transferred to between
old and young radiations. Such a presumed scenario causes the quantum field to
be singular at the horizon and hence a \textquotedblleft
firewall\textquotedblright\ \cite{amps13}, whose crossing by a free-falling
observer consequently leads to significantly different experience from the
nominal case predicted by general relativity. Like \textquotedblleft information loss paradox\textquotedblright\ ,
this \textquotedblleft firewall paradox\textquotedblright\ also results from
the conflict between quantum mechanics and general relativity and it
constitutes a renewed challenge.

If information can find a way to reside inside a black hole after Page's time,
the firewall could be evaded altogether. This brings up several earlier
suggestions involving remnants \cite{acn87}, which help to resolve the
information loss paradox if they possessed infinite degrees of freedom.
Quantum gravity does allow for a remnant, but its existence needs to be
substantiated (more discussions can be found in the recent review
\cite{coy15}). Would a remnant of infinite degrees of freedom imply a
satisfactory resolution for the information loss paradox? The answer, in fact,
is uncertain, basically because it remains unclear how to maintain
entanglement between the interior and the exterior of a black hole after
Page's time, unless the black hole itself has infinite degrees of freedom in
the interior regardless of its size. In particular, if a black hole with
finite degrees of freedom evolves into a remnant with infinite degrees of
freedom, how could this embody with unitarity?

This paper addresses the above controversy based on our recent results
\cite{zy17} that the volume of a noncommutative black hole is infinite. It is
organized as follows. First, we will revisit the firewall paradox and discuss
its intrinsic logic and the crucial elements in the second section. Then, we
describe an understanding for the black hole interior that is compatible with
the firewall proposal in the third section. In the forth section, we revisit
the infinite volume result for a noncommutative black hole. In the fifth
section, the resolution to the firewall paradox is presented based on the
renewed understanding for the black hole interior with noncommutative space.
Finally, we conclude in the sixth section. Throughout this paper, we use units
with $G=c=\hbar=k_{B}=1$.

\section{Information Loss Paradox and Firewall Paradox}

Firewall at horizon was first proposed in Ref. \cite{amps13} after the
discovery of inconsistency with black hole complementarity \cite{stu93} which
states that information is not lost and no other exotic phenomena occurs
within a set of postulates \cite{amps13}:

\begin{itemize}
\item[P1] Unitarity of black hole evaporation: The process of formation and
evaporation of a black hole, as viewed by a distant observer, can be described
entirely within the context of standard quantum theory. In particular, there
exists a unitary S-matrix which describes quantum theory.

\item[P2] Validity of effective field theory: Outside the stretched horizon of
a massive black hole, physics can be described to a good approximation by a
set of semi-classical field equations.

\item[P3] Existence of microscopic black hole entropy: To a distant observer,
a black hole appears to be a quantum system with discrete energy levels. The
dimension of the subspace of states describing a black hole of mass M is the
exponential of the Bekenstein entropy S(M).

\item[P4] Equivalence principle: A freely falling observer experiences nothing
out of the ordinary when crossing horizon.
\end{itemize}

First we discuss the inherent logic about these postulates. Hawking radiations
are thermal, which is almost a prerequisite to investigate any problems
already studied about Hawking radiations, unless the original calculation made
by Hawking are found erroneous. P2 provides a natural interpretation for the
spectrum due to entanglement between the outgoing Hawking mode and its
interior partner mode, as a ubiquitous feature of quantum field theory. This
does not lead to any conflict between the thermally distributed Hawking
radiations and unitarity of quantum mechanics. But P3 states that the
dimension of Hilbert space describing the radiation process of a black hole
can be regarded as exp[S(M)]. Thus, when half of the initial black hole
entropy is emitted away, the interior of the remaining black hole would not
have sufficient degrees of freedom to entangle with the outgoing radiations.
This indicates that the entanglement mechanism cannot ensure simultaneously
thermal radiations and unitarity of the radiation process after the Page's
time. If the sequentially emitted radiations were thermal, unitarity must be
violated unless a new mechanism is found. For the present situation, unitarity
or P1 requires that entanglement between the later outgoing Hawking mode $B$
and its interior partner mode $C$ is transferred to that between the early
outgoing Hawking mode $A$ and $B$. Entanglement between $B$ and $C$ could
ensure that Hawking radiations are thermally distributed, while the
transferred entanglement $AB$ preserves the unitarity of quantum mechanics. A
subtle point that is missing concerns how this transfer of entanglement takes
place. Black hole complementarity tells us that such a description does not
contradict with any fundamental physical principle since not a single observer
can see simultaneously the two types of entanglement mentioned above.

However, Almheiri \textit{et al }(AMPS) \cite{amps13} found that the
postulates P1, P2, P3 are inconsistent with P4, since they lead to the
violation of strong subadditivity of the entropy,%
\begin{equation}
S_{AB}+S_{BC}\geqslant S_{B}+S_{ABC}.
\end{equation}
The thermally distributed Hawking radiations satisfy $S_{B}>0$. Thus, this
violation can be derived by making use of the expression $S_{A}\geqslant
S_{B}+S_{A}$ because the first three postulates require $S_{AB}\leqslant
S_{A}$ but P4 implies $S_{BC}=0$ which is equivalent to the relation
$S_{ABC}=S_{A}$. In order to get rid of the contradiction, AMPS suggested that
P4 should be given up, which means that free infalling observers would not
feel the usual thing when crossing the horizon. The unusual thing called as
firewall challenges our present understanding of black hole physics and leads
to the breakdown of general relativity. As such it does not constitute a nice
resolution to black hole information loss paradox, but erects a new paradox,
called as firewall paradox, instead.

It is easily seen that the firewall paradox arises mainly from an incomplete
understanding of the black hole interior \cite{jp16}. Some proposals such as
those involving bulk reconstruction showed no firewall, i.e. using the
state-dependent operators in the black hole interior \cite{pr13,mp16},
building a connection between entanglement and Einstein-Rosen bridge (ER=EPR)
\cite{ms13}. In this paper, we describe a new understanding for black hole
interior with noncommutative space and investigate whether it can avoid the
firewall or not.

\section{Black Hole Interior}

In this section we will discuss the possibility of interpreting the black hole
entropy statistically by the degrees of freedom in the black hole interior.
According to the definition by Christodoulou and Rovelli (CR) \cite{cr15}, the
volume for a collapsed black hole is given by the maximal space-like
hypersurface $\Sigma$ bounded by a given surface $S$. For collapsed matter
described by the Eddington-Finkelstein coordinates
\begin{equation}
ds^{2}=-f(r)dv^{2}+2dvdr+r^{2}d\varphi^{2}+r^{2}\sin^{2}\varphi d\phi
^{2},\label{ef}%
\end{equation}
with $f(r)=1-{2M}/{r}$, the CR volume at late time is found to be
$V_{\mathrm{CR}}\sim3\sqrt{3}\pi M^{2}v\label{mv},$ for a Schwarzschild black
hole with an event horizon at $r=2M$. $v=t+\int{dr}/{f(r)}=t+r+2M\ln\left\vert
r-2M\right\vert $ denotes the advanced time. The maximal hypersurface occurs
at $r={3}M/{2}$ according to the method of auxiliary manifold \cite{cr15} or
maximal slicing of mathematical relativity \cite{bz15}. The volume
$V_{\mathrm{CR}}$ depends on the future behavior, e.g. evaporation, as it is
slated to the advanced time. Furthermore, it is worth noting that the value of
$V_{\mathrm{CR}}$ can be considerably larger than the observable universe for
a stellar black hole.

So how many degrees of freedom can be associated with such a large interior
volume? One can count the number of modes for a quantum field, e.g., a
massless scalar field $\Phi$ in the spacetime with the coordinates
\begin{equation}
ds^{2}=-dt_{w}^{2}+[-f(r)\dot{v}^{2}+2\dot{v}\dot{r}]d\lambda^{2}%
+r^{2}d\varphi^{2}+r^{2}\sin^{2}\varphi d\phi^{2},
\end{equation}
which results from Eq. (\ref{ef}) by the transformation $dv=\frac{-1}%
{\sqrt{-f}}\,dt_{w}+d\lambda$ and $dr=\sqrt{-f}\,dt_{w}$, and the dot on top
of a variable represents partial derivative with regard to the reparameterized
coordinate $\lambda$. The counting is carried out in the phase-space
\cite{quan} labeled by the positions $\{\lambda,\varphi,\phi\}$ and their
conjugate momenta $\{p_{\lambda},p_{\varphi},p_{\phi}\}$, in the space-like
hypersurface at $v>\!>M$ and $r={3}M/{2}$, and for the scalar field $\Phi$ in
the volume $V_{\mathrm{CR}}$. A single quantum state corresponds statistically
to a unit \textquotedblleft cell\textquotedblright\ of phase-space volume
$(2\pi)^{3}$ and the total number of quantum states arises after integrating
${d\lambda d\varphi d\phi dp_{\lambda}dp_{\varphi}dp_{\phi}}/{\left(
2\pi\right)  ^{3}}$ over the complete phase space, which gives the entropy
associated with $V_{\mathrm{CR}}$ as,%
\begin{equation}
S_{\mathrm{CR}}=\frac{\pi^{2}V_{\mathrm{CR}}}{45\beta^{3}}, \label{cre}%
\end{equation}
for the scalar field \cite{bz15,bz17}. It resembles the usual relationship
between entropy and volume, except that $V_{\mathrm{CR}}$ appears instead of
the usual volume.

In order to estimate the entropy $S_{\mathrm{CR}}$, we calculate the
parameters involved. First, a Schwarzschild black hole radiates and disappears
at $v\sim M^{3}$ according to the Stefan-Boltzmann law, which gives a large
but finite volume $V_{\mathrm{CR}}\sim$ $M^{5}$. Then, the temperature in the
black hole interior has to be worked out, which is not an easy task. But a
logical assumption is to take the value of the temperature which can equates
the entropy associated with the volume to the Hawking-Bekenstein entropy,
$S_{\mathrm{CR}}=S_{\mathrm{H}}=\frac{A}{4}$. This caters to the original
anticipation that the interior should have enough degrees of freedom to
interpret statistically the black hole entropy. A straightforward calculation
gives the required temperature $T=\beta^{-1}=\left(  \frac{\sqrt{3}\pi^{2}%
}{180}\right)  ^{-1/3}/M\simeq2.19M^{-1}$ which is much larger than the black
hole parameter $T_{\mathrm{H}}=1/8\pi M$.

Irrespective whether such an interior temperature is reasonable or not, it
leads to the statistical interpretation for the black hole entropy. But is
such an interpretation feasible? In order for Hawking radiations to remain
thermally distributed, these interior degrees of freedom must be entangled
with the outgoing radiations. When the black hole evaporates half of its
initial entropy away, the interior would not have enough degrees of freedom to
entangle with the subsequent outgoing radiations anymore. Therefore, either
the thermal spectrum has to be broken down or this entanglement is transferred
to other places in order to preserve unitarity. Coincidentally, this is just
the logic presented in the previous section through the postulates P1, P2 and
P3. Consequently, it also conflicts with P4 and leads to the so-called
firewall at the horizon. So this understanding for the black hole interior
cannot resolve the firewall paradox.

\section{Infinite Volume}

An alternative identification for the interior temperature is to take the
temperature $T=T_{\mathrm{H}}=1/8\pi M$, and we then find Eq. (\ref{cre})
reduces to
\begin{equation}
S_{\mathrm{CR}}\sim\frac{3\sqrt{3}\,}{45\times8^{3}}M^{2}=\frac{3\sqrt{3}%
\,}{90\times8^{4}\pi}A, \label{scr5}%
\end{equation}
with $A=16\pi M^{2}$ the area of event horizon. This states irrevocably that
entropy based on counting the modes of a scalar quantum field in
$V_{\mathrm{CR}}$ is proportional to the horizon \textit{surface} area that
binds the interior \textit{volume} \cite{bz15}. Unfortunately, it remains
short of providing sufficient number of modes to house all the information
because the prefactor in front of $A$ is much smaller than ${1}/{4}$ for the
Bekenstein-Hawking entropy.

The above estimate makes a critical assumption: the interior massless modes
are counted at a temperature equal to that of the horizon, which is plausible
if the interior and the exterior of a black hole are in thermal equilibrium.
This can be confirmed by the following discussion of the black hole
thermodynamics. When the volume of a black hole is taken into account,
thermodynamics call for a $PdV_{\mathrm{CR}}$ like term in the free energy,
whose presence would break the well established first law of black hole
thermodynamics. Thanks to the entropy associated with the volume, the term
$PdV_{\mathrm{CR}}\sim10^{-5}dM$ is nicely balanced out by the
$TdS_{\mathrm{CR}}\sim10^{-5}dM$ term. Collecting all terms together, the
first law of black hole thermodynamics becomes $dM=TdS-PdV_{\mathrm{CR}}$,
where $S=S_{\mathrm{H}}+S_{\mathrm{CR}}$ contains both a \textit{surface
}$S_{\mathrm{H}}$ and a \textit{volume }$S_{\mathrm{CR}}$ contribution. So
what is the meaning for the pressure $P$ introduced above? It can be
understood as arising from vacuum polarization \cite{fn98} due to zero point
fluctuations of the local energy density, which gives a quantum pressure
$P=1/{\left(  90\times8^{4}\pi^{2}{M^{4}}\right)  }$ at the horizon
\cite{wgu76,pc80,dnp82,te83}. Thus, the interior temperature taken here is
justified and self-consistent with black hole thermodynamics.

In this perspective, more or even infinite degrees of freedom inside a black
hole are needed in order to resolve the information loss paradox. Several
progresses have been made along this direction as well. First, backaction can
be included when Hawking radiation is considered. The dynamic evolution for a
spherically symmetric black hole due to Hawking radiation can be described by
the well-known Vaidya metric \cite{pv51}, with which $V_{\mathrm{CR}}$ is
calculated and the changes are however, found to be insignificant
\cite{yco15,cl16} if the advanced time $v$ is estimated by using the
Stefan--Boltzmann law. The entropy associated with $V_{\mathrm{CR}}$ including
backaction is also worked out and the modification to Eq. (\ref{scr5}) is
found to be negligible as well \cite{zy17}.

Thus, other reasons must be found to explain the insufficient counting number
(\ref{scr5}). For instance, it could be due to the inaccurate estimate for $v$
based on the Stefan-Boltzmann law, which holds only for a black hole with a
mass much greater than the Planck mass \cite{sm95}. An improved treatment for
the final evaporation stage can come from noncommutative space. In analogy to
canonical conjugate variables in quantum mechanics, when physics enter into
the Planck scale, spacetime coordinates become noncommutative, e.g., $\left[
x^{\mu},x^{\nu}\right]  =i\theta\epsilon^{\mu\nu}$ where the parameter
$\theta$ is of the dimension length squared. First proposed in the 40s of the
last century \cite{hs47}, the idea of noncommutative spacetime has since been
extensively applied to black hole physics (see the review \cite{pn09} and
references therein). Its direct application, however, is rather inconvenient.
An alternative method was provided \cite{nss06} which attributes the spatial
noncommutative effect to a modified energy-momentum tensor, or a source while
the Einstein tensor remains unchanged \cite{mkp07}. It is consequently found
that the final evaporation stage can be well described by a remnant of mass
$M_{r}\neq0$, and the singular divergent temperature of a commutative
Schwarzschild black hole is avoided.

For noncommutative space, the Eddington-Finkelstein coordinates are revised
into $f(r)=1-\frac{4M}{r\sqrt{\pi}}\gamma\left(  \frac{3}{2},\frac{r^{2}%
}{4\theta}\right)  $, where $\gamma(\mu,z)$ denotes the lower incomplete gamma
function. The modified spacetime structure dominates only near the Planck
scale, but is generally negligible for a black hole of large mass. As a
result, the corresponding thermodynamic parameters take their approximate
(commutative) forms, including the expressions for $V_{\mathrm{CR}}$ and
$S_{\mathrm{CR}}$, before approaching the Planck scale. In the final
evaporation stage as $M$ approaches $M_{r}$, the temperature is approximately
proportional to the mass instead of the usual inversely-proportional
relationship \cite{bch73}. The advanced time becomes
\begin{equation}
v\simeq\frac{1}{\left(  M-M_{r}\right)  ^{3}}\,,
\end{equation}
in the limit of large $v$ \cite{mkp07}. The final evaporation stage thus needs
an infinitely long time to complete, although the overall change to the black
hole size will only be several $\sqrt{\theta}$. The volume associated
statistical entropy formally remains the same as in Eq. (\ref{cre}) for
commutative space, except for the modification to the volume itself
\cite{zy17}
\begin{equation}
V_{\mathrm{NCR}}\sim\frac{M^{2}}{\left(  M-M_{r}\right)  ^{3}}, \label{ncref}%
\end{equation}
as $v\rightarrow\infty$. This describes a black hole with diverging volume
($M\rightarrow M_{r}$) wrapped by a finite horizon. The associated entropy,
$S_{\mathrm{NCR}}$, also diverges, although at a slower rate. A noncommutative
black hole thus possesses larger information storage capacity, which indicates
that its statistical interpretation for the Bekenstein-Hawking entropy must be
independent of the interior \cite{jmr05,mkp06,hs10}. A further refinement
could include the early emission stage before Planck scale is reached.
However, this change is insignificant because the estimated result from the
final stage alone is already divergent.

Thus, we have presented a new scenario for the black hole interior resulting
in a divergent volume. In the next section we will study whether and how the
firewall can be avoided based on this result.

\section{Possible Resolution to Firewall Paradox}

As stated in the second section, a crucial element for the proposal of
firewall at horizon is the requirement of transferring entanglement between
the outgoing Hawking radiation and its partner inside the hole to entanglement
between early and later outgoing radiations. The postulate P3 indicates that
the black hole interior would not have enough degrees of freedom to maintain
entanglement with outgoing radiations after half of the initial entropy is
emitted. However, if the black hole interior is large enough to house the
required degrees of freedom during the complete radiation process, this
transfer of entanglement will not be needed, which means that the firewall
will be avoided.

For the situation considered in this paper, since the entropy associated
volume, $S_{\mathrm{NCR}}$, is divergent, at any time of evaporation a
noncommutative black hole thus always has enough degrees of freedom to
entangle with outgoing Hawking radiations. This guarantees the radiations can
be thermally distributed. Hence, the description about unitary evolution of
black hole radiation can be changed to accommodate unitarity by assuming that
information resides inside a black hole during the whole Hawking radiation
process without the requirement of entanglement transfer, which hence
precludes the firewall.

A brief description for the unitary process without firewall can be made
accordingly by considering a black hole in the process of emitting radiations.
Its complete state constituting of the interior $\left\vert \Psi_{k}%
^{(i)}\left(  \varphi\right)  \right\rangle $ and the exterior $\left\vert
\Psi_{k}^{(e)}\right\rangle $ (Hawking radiations) can be expressed formally
as below on the right-hand-side
\begin{equation}
\left\vert \varphi\right\rangle \rightarrow\mathrm{evolve}\rightarrow\sum
_{k}c_{k}\left\vert \Psi_{k}^{(i)}\left(  \varphi\right)  \right\rangle
\left\vert \Psi_{k}^{(e)}\right\rangle , \label{uni}%
\end{equation}
in an entangled form, which in the end leaves Hawking radiations in a mixed
state after the initial state $\left\vert \varphi\right\rangle $ composed of
matter is collapsed into a black hole and evaporates. Of course, this scenario
depends critically on whether the interior of a black hole possesses
sufficient space to house information.

For the above scenario to work, the final stage of unitary evolution
(\ref{uni}) has to be treated well. As stated in the last section, spatial
noncommutativity gives a plausible description for the final stage of black
hole evaporation, which provides a remnant as the ending point of evaporation.
In particular, the result discussed above that a noncommutative black hole
contains divergent degrees of freedom reconciles the previously discussed
remnant-based resolution \cite{coy15} for information loss paradox,
irrespective of whether the remnants possess infinite degrees of freedom or
not. A remnant possessing infinite degrees of freedom constitutes a happy
ending to unitary evolution for a black hole with infinite degrees of freedom
to begin with. There is no need to worry about a remnant with finite degrees
of freedom because such a state cannot be reached within a finite time as long
as the black hole possesses infinite degrees of freedom. In addition, if a
remnant were thrown into a black hole possessing infinite degrees of freedom,
no confusion or conflict with unitarity or thermodynamics will resurrect.
Information loss thus do not happen for a black hole with an infinite
interior. Unitary is assured and no firewall is needed.

\section{Conclusion}

In conclusion, the volume of a black hole is discussed for its potential and
significant role in resolving black hole information loss paradox. When a
black hole forms and starts to evaporate, unitary dynamics respecting
Hawking's thermal radiation emissions are understood in terms of entanglement
between the internal degrees of freedom and the outgoing radiations of a black
hole. A plausible approach to evade firewall assumes that information is
stored inside a black hole. Its validity requires the interior of a black hole
to possess a sufficiently large volume capable of maintaining entanglement
during the whole evaporation process. Our study suggests the semiclassical
result for the interior volume is insufficient even for the large internal
volume obtained according to the CR definition. When quantum gravity effect is
taken into account using noncommutative space, the treatment for the final
evaporation stage can be improved, leading to a divergent volume capable of
housing infinite degrees of freedom. This presents a suggestive example to
realize unitary evolution with entanglement based mechanism that calls for no firewall.

B. C. Zhang thanks Qingyu Cai for his helpful comments. This work is supported
by the NSFC (grant No. 11654001, No. 11747605 and No. 91636213).

\end{document}